\documentstyle[prl,psfig,aps,multicol]{revtex}

\psfigurepath{/month/saif/Part2/latex/}

\begin{document}
\draft
\title{Quantum Recurrences in Periodically Driven Systems}
\author{Farhan Saif}
\address{Department of Electronics, Quaid-i-Azam University, Islamabad, Pakistan.}
\date{\today}
\maketitle

\begin{abstract}
We investigate the quantum recurrence phenomena in periodically driven systems.
We calculate the classical period and the quantum recurrence time and develop their
interdependence. We further
predict the behavior of the recurrence phenomena for the power law potentials.
\end{abstract}

\maketitle

\begin{multicols}{2}
\narrowtext

\section{Introduction}
\label{dynrev}

The phenomena of recurrence or revival is a beautiful combination of classical mechanics,
wave mechanics, and quantum laws.
A wave packet evolves over a short period of time
in a potential, initially following classical mechanics. It spreads while moving along its
classical trajectory, however rebuilds itself after a classical period. It follows
wave mechanics in its long time
evolution and gradually observes a collapse.
However, the discreteness of quantum mechanics leads to the restoration and
restructuring of the wave packet.

In one degree of freedom systems the phenomena of quantum revivals are well studied both
theoretically and experimentally. The quantum revivals were first studied in
cavity quantum electrodynamics~\cite{kn:eber,kn:naro,kn:yurke}. Recently,
the existence of revivals has been investigated in atomic
~\cite{kn:park,kn:alber,kn:aver,kn:alber1,kn:brau,kn:leic,kn:leic1,kn:stroud}
and molecular~\cite{kn:greb,kn:fisc,kn:vrak,kn:donch} wave packet evolution.

The periodically driven quantum systems~\cite
{kn:haak,kn:hogg,kn:bres}, and two-degree-of-freedom systems such as stadium
billiard~\cite{kn:toms} indicate the presence of quantum revivals in higher
dimensional systems. Latter, it is proved that the recurrence or revival phenomenon
is a generic property of the one degree of freedom periodically driven
quantum systems~\cite{saifm}. In the present contribution we calculate the
classical period and quantum revival time for the driven systems. Moreover,
we calculate their interdependence for different dynamical regimes. Latter
we explain these interdependences for the power law potentials.

The layout of the paper is as follows: In Sec. II, we write general Hamiltonian for the
periodically driven time dependent systems.
In Sec. III, we calculate the quasi-energy eigen functions
and quasi-energies for these systems. In Sec. IV we calculate the
classical period and quantum recurrence time and write their interdependence in
Sec. V. We dedicate section VI for a discussion of this interdependence in
power law potentials.


\section{Periodically Driven Systems}

In its evolution in time a material wave packet displays quantum revivals
in a one dimensional system driven by an external periodic
force~\cite{saifm}.
Subject to the strength of the external modulation,
the classical dynamics may support stable and unstable motion.
The Hamiltonian of the system in the dimensionless form is expressed
as,
\begin{equation}
H=H_0+ \lambda\,V(z)\sin t\;.
\label{eq:sche}
\end{equation}
Here, $H_0=p^2/2m+V_1$, is the Hamiltonian of the undriven system, and $V_1$ is the
potential of the undriven system.
The eigen states and the eigen values of the time independent
system are, $|n\rangle$, and, $E_n$, respectively.
Moreover, $\lambda$ expresses the dimensionless modulation strength, and $V(z)$
defines the coupling between the one dimensional potential and the external
modulation.

The measurement of the time of the revival in the classical domain and in the quantum domain
requires the solution of the periodically driven quantum system.
In the time dependent system the energy is no more a
constant of motion. Therefore, the periodically driven system is made
to solve in the region of resonances by using the secular perturbation theory~\cite{kn:born}.
In this approach, the faster frequencies are averaged
out and the dynamical system is effectively reduced to
one degree of freedom, which is integrable.
Thus, the solution of the periodically driven system yields
the quasi-eigen energy and the quasi-energy eigen states.


\section{Quasi-Energy and Quasi-energy Eigen States}

In order to calculate the quasi- energy eigen states and the quasi-energy of the
driven system, we~\cite{Berman1,kn:flat}
make the ansatz that the solution of the
Schr\"odinger equation corresponding to the Hamiltonian~(\ref{eq:sche}) is
written as,
\begin{equation}
|\psi(t)\rangle=\sum_{n} C_n(t) |n\rangle
\exp\left\{-i\left[E_r+(n-r)\frac{\hbar}{N}\right]
\frac{ t}{\hbar }\right\}\;,
\label{eq:1}
\end{equation}
where, $\hbar$ is the scaled Planck's constant,
and $E_r$ is the average energy of the wave-packet in the $N$th resonance.

We substitute Eq.~(\ref{eq:1}) in the time dependent Schr\"odinger equation. By
following the method of secular perturbation
theory, we~\cite{kn:flat} get the transformed Schr\"odinger
equation as,
\begin{equation}
i\hbar\frac{\partial g}{\partial t}=
\left[-\frac{{\hbar}^2 N^2 \zeta}{2}\frac{\partial^2}{\partial\theta^2}
-iN\hbar\left(\omega-\frac{1}{N}\right)\frac{\partial}{\partial\theta}
+H_0 +\lambda V\sin(\theta)\right]g\;.
\label{eq:5}
\end{equation}
The scaled parameters, $\zeta=E''_r/\hbar^2$, and, $\omega=E'_r/\hbar$, physically
express the nonlinearity in the system and its frequency, respectively.
Here, $E''_r$, defines the second
derivative and $E'_r$ defines the first derivative of the
unperturbed energy, calculated at the mean quantum number, $n=r$.

We apply the method of factorization, such that,
\begin{equation}
g(\theta,t)={\tilde g}(\theta)e^{-i{\cal E}t/\hbar}e^{-i(\omega-1/N)\theta/N\zeta\hbar}
\end{equation}
and change the variable, $\theta$, as $\theta=2z+\pi/2$. The substitutions reduce the
Eq.~(\ref{eq:5}) to the
standard Mathieu equation~\cite{kn:abra}, viz.,
\begin{equation}
\left[\frac{\partial^2}{\partial z^2}
+a-2q\cos(2 z)\right]
\tilde{g}(z)=0\;.
\label{eq:6}
\end{equation}
Here,
\begin{equation}
a=\frac{8}{N^2\zeta\hbar^2}\left({\cal E}-{\bar H}_0+\frac{(\omega-1/N)^2}{2\zeta}\right),
\end{equation}
where, ${\cal E}$ is the quasi-energy of the system.
Moreover, $q=4\lambda V/N^2\zeta\hbar^2$, and
$\tilde{g}(z)$ is a $\pi$-periodic function.

The function $\tilde{g}(\theta)$ is related to $C_n(t)$ as,
\begin{eqnarray}
C_m&=&\frac{1}{2\pi}\int\limits_{0}^{2\pi} g(\varphi)
e^{-i(m-r)\varphi}\,d\varphi \nonumber\\
&=&\frac{1}{2N\pi}\int\limits_{0}^{2N\pi} g(\theta)
e^{-i(m-r)\theta/N}\,d\theta\;.
\label{eq:3a}
\end{eqnarray}

Hence, the Floquet quasi-energy eigen functions are
\begin{equation}
|\psi_k(t)\rangle=e^{-i{\cal E}_k
t/\hbar}\,|u_k(t)\rangle\;,
\label{eq:qv}
\end{equation}
where, quasi energy eigen values ${\cal E}_k$, and $|u_k(t)\rangle$ are defined as
\begin{eqnarray}
{\cal E}_k\equiv\frac{\hbar^2N^2\zeta}{8}a_{\nu(k)}-
\frac{(\omega-1/N)^2}{2\zeta}+{\bar H}_0\;,
\label{eq:eneg} \\
|u_k(t)\rangle\equiv\frac{1}{2\pi}\sum_n
e^{int/N}\int\limits_{0}^{2\pi}
d\varphi e^{i\nu N\varphi/2} e^{-i(n-r)\varphi/2}
P_{\nu(k)}\,|n\rangle \;.
\label{eq:amp}
\end{eqnarray}
In this way, we have obtained an approximate solution for a nonlinear
resonance of our explicitly time-dependent system.

In order to obtain a $\pi$-periodic solution in $\varphi$-coordinate, we
require the coefficient of $\varphi$ in the exponential factor to be equivalent
to an even integer, $k$. This condition provides
the value for the index $\nu$ as,
\begin{eqnarray}
\nu=\frac{2k}{N}+\frac{2(\omega-1/N)}{N\zeta\hbar}.
\label{eq:denu}
\end{eqnarray}


\section{The Quantum Recurrence times}

An initial excitation produced at the action, $I=I_0$, observes various time scales at which it
reappears completely or partially during its evolution in the dynamical system.
In order to find these time scales
at which revivals occur in the quantum mechanical modulated system,
we employ the eigen energy ${\cal E}_k$ of the system~\cite{saifm,saifPLA}.

These time scales, $T_{\lambda}^{(j)}$,
are inversely proportional to the frequencies, $\omega^{(j)}$, where
$j$ is an integer. The frequency, $\omega^{(j)}$, is defined as
\begin{equation}
\omega^{(j)}=(j!)^{-1}\hbar^{(j-1)} \frac{\partial^{(j)}{\cal E}_k}{\partial I^{(j)} }
=(j!\hbar)^{-1} \frac{\partial^{(j)}{\cal E}_k}{\partial n^{(j)} },
\end{equation}
calculated at $I=I_0=n_0\hbar$. The index $j$ describes the order of
differentiation of the quasi energy, ${\cal E}_k$. With the increasing values of
$j$, we have smaller frequencies which indicate longer times for the higher-order revivals.

The time scale, $T_{\lambda}^{(1)}=T_{\lambda}^{(cl)}$, defines classical
period of the driven system and is inversely proportional to
$\omega^{(1)}$. In the absence of external periodic modulating force,
the frequency $\omega^{(1)}$ reduces to $\omega$.

The time scale, $T_{\lambda}^{(2)}=T_{\lambda}^{(Q)}$,
defines quantum mechanical revival time in the
modulated systems. It has inverse proportionality with $\omega^{(2)}$.
Here, we have
$\omega^{(2)}=(2!)^{-1}\hbar\partial^2 {\cal E}_k/\partial I^2|_{I=I_0}$.
Partial reappearance of the initially excited wave packet occurs, at the fractions of the
revival time. Therefore, it is appropriately named as fractional revival phenomenon.

On substituting the value for the quasi energy, ${\cal E}_k$, from Eq.~(\ref{eq:eneg})
in $\omega^{(1)}$, we obtain the classical period as,
\begin{equation}
T_{\lambda}^{(cl)}=[1-M_o^{(cl)}]T_0^{(cl)}\Delta.\label{eq:clt1}
\end{equation}
By making the same substitution
in, $\omega^{(2)}$, we find the quantum revival time
for the driven system as,
\begin{equation}
T_{\lambda}^{(Q)}=[1-M_o^{(Q)}]T_0^{(Q)}.\label{eq:clt2}
\end{equation}
Here, the time scales,
$T^{(cl)}_0(\equiv 2\pi/\omega)$, and
$T^{(Q)}_0 (\equiv 2\pi(\frac{\hbar}{2!}\zeta)^{-1})$, define
the classical period and
the quantum revival time in the absence of external modulation. Moreover,
$\Delta=(1-\omega_N/\omega)^{-1}$ where $\omega_N=1/N$.

The time modification factor
$M_o^{(cl)}$ and
$M_o^{(Q)}$ are given as,
\begin{equation}
M_o^{(cl)}=-\frac{1}{2}\left(\frac{\lambda V \zeta\Delta^2} {%
\omega^2} \right)^2 \frac{1}{(1-\mu^2)^2}\, ,
\label{eq:modf1}
\end{equation}
and
\begin{equation}
M_o^{(Q)}=\frac{1}{2}\left(\frac{\lambda V \zeta\Delta^2} {%
\omega^2} \right)^2 \frac{3+\mu^2}{(1-\mu^2)^3}%
\,  \label{eq:modf2}
\end{equation}
where,
\begin{eqnarray}
\mu=\frac{N^2\hbar\zeta\Delta}{2\omega}.
\label{mu}
\end{eqnarray}
Equations
~(\ref{eq:clt1}) and
~(\ref{eq:clt2}) express
the classical period and
the quantum revival time in a one degree of freedom system in
the presence of an external modulation.
These time scales are function of the modulation strength $\lambda$,
the frequency, $\omega$, and the nonlinearity, $\zeta$, associated with the unmodulated
system.

As the modulation term vanishes, that is $\lambda=0$, the modification terms
$M_o^{(cl)}$ and $M_o^{(Q)}$ disappear. Thus,
from Eqs.~(\ref{eq:clt1}) and (\ref{eq:clt2}),
it is obvious that the classical period and the quantum revival time in the
presence and in the absence of external modulation are equal, that is,
$T_{\lambda}^{(cl)}=T_0^{(cl)}$ and $T_{\lambda}^{(Q)}=T_0^{(Q)}$.
As there exist no resonances for $\lambda=0$, we find $\Delta=1$.


\section{Classical period and quantum revival time: Interdependence}

The nonlinearity present in the energy spectrum of the undriven system, contributes
to the classical period and the quantum revival time in the presence and in the
absence of an external modulating force~\cite{saifPLA}.

\subsection{Vanishing nonlinearity}

In the absence of nonlinearity in the energy spectrum, {\it i.e} for $\zeta=0$,
the time modification factor for the classical period
$M_o^{(cl)}$ and for the quantum revival time $M_o^{(Q)}$ vanish,
which is evident from Eqs.~(\ref{eq:modf1}) and (\ref{eq:modf2}).
Thus, a periodically driven linear system displays the quantum revivals
after infinite time, {\it i.e} $T_{\lambda}^{(Q)}= T_0^{(Q)}=\infty$.

Hence, in the modulated linear
system only classical period exists.
The system displays
revivals after the classical period, that is
$T_{\lambda}^{(cl)}=T_0^{(cl)}\Delta=2\pi\Delta/\omega$.


\subsection{Weak nonlinearity}

For weakly nonlinear energy spectrum, the classical period, $T_{\lambda}^{(cl)}$, and the
quantum revival time, $T_{\lambda}^{(Q)}$, in the presence of modulation are
related with, $T_0^{(cl)}$, and, $T_0^{(Q)}$, of the unmodulated system as,
\begin{equation}
3T_{\lambda}^{(cl)}T_0^{(Q)}+\Delta T_0^{(cl)}T_{\lambda}^{(Q)}=4\Delta T_0^{(Q)}T_{0}^{(cl)}.
\label{eq:r1}
\end{equation}

Since the quantum revival time, $T_{\lambda}^{(Q)}$,
and, $T_{0}^{(Q)}$, depend inversely on nonlinearity in the unmodulated system,
they are much larger than the classical period, $T_{\lambda}^{(cl)}$, and $T_0^{(cl)}$.
Thus, the time modification factors $M_o^{(cl)}$ and $M_o^{(Q)}$ are related as,
\begin{equation}
M_o^{(Q)}=-3 M_o^{(cl)}=-3\alpha,
\label{moq}
\end{equation}
where
\begin{equation}
\alpha= \frac{1}{2}\left(\frac{\lambda V \zeta}{{\omega}^2}\right)^2.
\label{alphaq}
\end{equation}
The modification factor $\alpha$ is directly proportional to the
square of the nonlinearity, $\zeta^2$ in the
system.

In the asymptotic limit, {\it i.e.} for $\zeta$ approaching zero,
the quantum revival time in the modulated and unmodulated system are equal and infinite,
that is $T_{\lambda}^{(Q)}=T_0^{(Q)}=\infty$. Furthermore, the classical period in the
modulated and unmodulated cases are related as,
$T_{\lambda}^{(cl)}=T_0^{(cl)}\Delta$, as mentioned above.

The quantum revival time $T_{\lambda}^{(Q)}$ reduces
by $3\alpha T_0^{(Q)}$, whereas, $T_{\lambda}^{(cl)}$ increases by
$\alpha T_0^{(cl)}$.


\subsection{Strong nonlinearity}

For strongly nonlinear energy spectrum, the classical period, $T_{\lambda}^{(cl)}$, and the
quantum revival time, $T_{\lambda}^{(Q)},$ in the presence of the coupling are
related with the, $T_0^{(cl)}$, and, $T_0^{(Q)},$ of the uncoupled system as,
\begin{equation}
T_{\lambda}^{(cl)}T_0^{(Q)}-\Delta T_0^{(cl)}T_{\lambda}^{(Q)}=0.
\label{eq:r2}
\end{equation}

We have the time modification factors related as
\begin{equation}
M_o^{(Q)}=M_o^{(cl)}=-\beta,
\label{relm}
\end{equation}
where
\begin{equation}
\beta= \frac{1}{2}\left(\frac{4\lambda V}{N^2 \zeta \hbar^2}\right)^2.
\label{betaq}
\end{equation}

Hence, for a highly nonlinear case $\beta$ approaches to zero. Thus,
the time modification factors,
given in Eqs.~(\ref{eq:modf1}) and (\ref{eq:modf2}),
vanish both in the classical domain and in the quantum domain.
As a result, the Eqs.~(\ref{eq:clt1}) and (\ref{eq:clt2}) reduces to,
$T_{\lambda}^{(cl)}=T_{0}^{(cl)}\Delta$, and
$T_{\lambda}^{(Q)}=T_{0}^{(Q)}$, which proves the equality given in Eq.~(\ref{eq:r2}).

The quantity $\beta$ which determines the
modification both in classical period and in quantum revival time is inversely depending on
fourth power of Planck's constant $\hbar$, hence for highly quantum mechanical cases we find
that the revival times remain unchanged.

\section{Quantum Recurrences and Power law potentials}

We may express a large number of potentials as power law potentials~\cite{Robinett}, such that
$V_1(z)=V_0|\frac{z}{a}|^k$, where $V_0$ and $a$ are arbitrary constants.
For the corresponding one degree of freedom system, the energy eigen value is
\begin{equation}
E_n^{(k)}=\left[\left(n+\frac{1}{2}\right) \frac{\pi\hbar}{2a\sqrt{2 m}}V_0^{1/k}
\frac{\Gamma(1/k+3/2)}{\Gamma(1/k+1)\Gamma(3/2)}\right]^{2k/k+2}.
\end{equation}

The corresponding classical frequency and nonlinearlity in the absence of any external force is
defined as
\begin{equation}
\omega=\frac{1}{\hbar}\frac{2k}{k+2}\left(r+\frac{1}{2}\right)^{-1}E_r^{(k)},
\end{equation}
and
\begin{equation}
\zeta=\frac{1}{\hbar^2}\frac{2k(k-2)}{(k+2)^2}\left(r+\frac{1}{2}\right)^{-2}E_r^{(k)}.
\end{equation}

Hence the classical period and the quantum recurrence time read as
\begin{equation}
T_0^{(cl)}=\frac{2\pi\hbar}{E_r^{(k)}}\frac{k+2}{2k}\left(r+\frac{1}{2}\right),
\label{tpl1}
\end{equation}
and
\begin{equation}
T_0^{(Q)}=\frac{4\pi\hbar}{E_r^{(k)}}\frac{(k+2)^2}{2k(k-2)}\left(r+\frac{1}{2}\right)^2,
\label{tpl2}
\end{equation}
respectively.

We substitute the values of $T_0^{(cl)}$ and $T_0^{(Q)}$ from Eqs.~(\ref{tpl1}) and
(\ref{tpl2}) in the Eqs.~(\ref{eq:r1}) and (\ref{eq:r2}). Thus for a weak nonlinearity
in the power law potentials, the interdependence between
the classical period and the quantum recurrence time in the presence of external periodic
modulation becomes
\begin{equation}
T_{\lambda}^{(Q)}=-3C_k T_{\lambda}^{(cl)}+4T_{0}^{(Q)},
\label{rp1}
\end{equation}
and, for a strong nonlinearity it reads as,
\begin{equation}
T_{\lambda}^{(Q)}=C_k T_{\lambda}^{(cl)},
\label{rp2}
\end{equation}
respectively. Here, we define $C_k$ as
\begin{equation}
C_k=\frac{2}{\Delta}\left(\frac{k+2}{k-2}\right)\left(r+\frac{1}{2}\right).
\end{equation}


\section{Acknowledgement}

I submit my thanks to Higher Education Commission, Pakistan, research grant for
the research work.

\end{multicols}

\end{document}